\documentstyle[aps]{revtex}

\input epsf

\begin{document}

\title{Symmetric Diblock Copolymers in Thin Films (II):\\
Comparison of Profiles between \\
Self-Consistent Field Calculations and Monte Carlo Simulations.
}

\author{
T. Geisinger, M.\ M\"{u}ller, and K.\ Binder
\\
{\small Institut f{\"u}r Physik, WA 331, Johannes Gutenberg Universit{\"a}t}
\\
{\small D-55099 Mainz, Germany}
}
\date{\today, draft to be submitted to J.Chem.Phys.}
\maketitle

\begin{abstract}
The structure of lamellar phases of symmetric $AB$ diblock copolymers in a thin film is investigated.
We quantitatively compare the composition profiles and profiles of individual segments in  
self-consistent field calculations with Monte Carlo simulations in the bond fluctuation model for chain length $N=32$
and $\chi N=30$.  Three film thicknesses are investigated, corresponding to parallel oriented lamellae with 2 and 4 interfaces and a 
perpendicular oriented morphology. Taking account of capillary waves, we find good quantitative agreement between 
the Monte Carlo simulations and the self-consistent field calculations. However, the fluctuations of the local interfacial
position are strongly suppressed by confinement and mutual interactions between lamellae.
\end{abstract}

\section{ Introduction. }
Confinement of spatially structured fluids into thin films gives rise to a rich and interesting interplay between the
intrinsic length scale of their structure in the bulk and the geometry of the film. One the one hand, confinement 
generally modifies the phase diagram of fluids\cite{Nakanishi}. For spatially structured phases the confinement leads to
transitions between phases with identical symmetry but different orientation with respect to the confining walls.
On the other hand the surfaces alter the local structure of the fluid 
in their vicinity. This gives rise to the enrichment of one component at the surface and the formation of wetting layers, but it
also entails changes in the local fluid structure (e.g., packing and alignment effects).  Controlling the properties of confined 
fluids is important for various practical applications. The orientation of the morphology is, for instance, important for
connecting the self-assembled material to other devices (e.g., to form a molecular sieve\cite{Russell}).

Studying effects of confinement on spatially structured polymeric fluids is particularly rewarding, because 
the large length scale of the phenomena, set by the chain extension, leads to a rather universal behavior, i.e., many features are 
independent from the details of the chemical architecture. The equilibrium properties of polymeric systems are usually well describable 
by self-consistent field calculations. Moreover, the large length scales facilitate the application of experimental techniques. The 
detailed composition profiles in the vicinity of surfaces and even profiles of individual segments are experimentally accessible. 
However, the large lengths scale of the phenomena goes along with protracted long time scales to reach equilibrium.
There are quantitative comparisons between experiments and theory: The profiles of individual segments in films of diblock copolymers and their blends
have been measured by small angle neutron scattering, and the results have been compared to self-consistent field calculations\cite{C_SULL,C_MATSEN}. The 
studies found that the self-consistent field calculations describe the qualitative features, but that the profiles were broadened by long wavelength
fluctuations of the local position of the interfaces. 

In the previous paper\cite{PAPER1}, hereafter denoted as paper I, we have used a self-consistent field technique similar to Matsen\cite{MSCF} for 
calculating the phase diagram of a symmetric $AB$ diblock copolymer melt confined into a thin film as a function of the film thickness 
and temperature. The calculations have been compared to Monte Carlo simulations in the framework of the bond fluctuation model at 
intermediate segregation ($\chi N=30$). Depending on the film thickness the systems assembled into parallel oriented lamellar phases
$L_2$ and $L_4$, in which $2$ or $4$ $AB$ interfaces run parallel to the surfaces of the film, or a perpendicular lamellar phase $L_\perp$
in which the interfaces run perpendicular to the confining surfaces. The dependence on the film thickness exhibited by the Monte Carlo 
simulation is in agreement with the self-consistent field calculations.

The aim of the present paper is to investigate the local structure of these films in more detail.
The detailed profiles of the composition and individual segments provide a sensitive testing bed  for comparing Monte Carlo simulations and self-consistent field theory.
While the comparisons between experiments and theory have provided many valuable insights, Monte Carlo simulations might contribute to our understanding by investigating 
model systems in which many different quantities are simultaneously accessible. In the simulations we can analyze profiles of different quantities on various lateral length 
scales, and we can compare the $AB$ diblock copolymer melt with a blend of $A$ and $B$ homopolymers of identical architecture.

There are several potential sources of deviations between the Monte Carlo simulations/experiments and the self-consistent field calculations: 
({\bf i}) At low incompatibility, i.e., around the onset of spatial ordering, fluctuations become important. These shift the transition temperature from its 
mean field value $\chi N = 10.5$ to higher values of the incompatibility, and the transition becomes first order\cite{Fredrickson}. Shifts of the
transition temperature of similar magnitude are predicted in the P-RISM framework by Schweizer and co--workers\cite{David}. In our Monte Carlo simulations we 
observe the order-disorder transition in the regime $13.5 \leq \chi N \leq 17$ (cf.\ Fig.7 in paper I). At higher segregation, however, fluctuation effects become less important.
({\bf ii}) The self-consistent field theory assumes the interface between the $A$ and $B$ domains to be ideally flat in the lamellar phase. However, there
are long wavelength fluctuations of the local interfacial position (i.e., capillary waves). Their free energy cost vanishes as their wavelength diverges, and, hence,
they are important at all temperatures. These fluctuations lead to a broadening of the laterally averaged profiles measured in experiments or Monte Carlo simulations.
({\bf iii}) Moreover, computer simulations\cite{Fried} and experiments\cite{Maurer,Stamm1,Stamm} reveal that the copolymers
assume a dumbbell--like shape already above the order-disorder transition temperature, as to avoid energetically unfavorable contacts between their different moieties. This 
leads to an increase of the chain extension. The interplay between the inter- and intramolecular energy has been studied in the P-RISM framework by Schweizer and co-workers\cite{David} 
and field theoretical approaches\cite{Vilgis}
({\bf iv}) At high incompatibility ($\chi \sim {\cal O}(1)$), the smallest length scale of the spatial structure (e.g., the width of the interface between the $A$ and $B$ domains)
decreases rapidly. If this length scale becomes comparable to the length scale of the underlying microscopic structure (e.g.,  persistence length of the polymer
or the range of packing effects of the monomers) a description in the framework of the Gaussian chain model breaks down. In fact, even for simple systems like
interfaces between polymers of different persistence length the Gaussian chain model gives qualitatively erroneous predictions at high incompatibility\cite{MW}. 
Density functional or P-RISM calculations of more realistic models might account for the changes in the local fluid structure. Comparing self-consistent field calculations
in the framework of the Gaussian chain model and density functional theory for a model with local fluid structure, Nath and co--workers\cite{Nath1,Nath2}
observed rather pronounced deviations between both approaches.
({\bf v}) The confining surfaces create a sharp density gradient. In this region the local structure of the polymeric fluid is important at all incompatibilities. The 
monomer density exhibits pronounced oscillations in the vicinity of the walls (cf.\ Fig.8 in paper I). Moreover, polymers align parallel to the wall. The latter effect
is partially captured in the self-consistent field calculations and favors a perpendicular arrangement of the lamellae\cite{Pickett,Sommer}.
The aim of our paper to present a quantitative comparison between the Monte Carlo simulations and self-consistent field calculations and to quantify the 
effects discussed above in the framework of a well-studied model.

The outline of our paper is as follows: In the next section we give a brief synopsis of the model and the computational technique.
The reader is refered to paper I for further technical details. Then we discuss the strength of interfacial fluctuations and compare the
laterally averaged profiles of the Monte Carlo simulations with those of the self consistent field calculations. Taking account
of interfacial fluctuations we find good agreement, but the strength of the fluctuation effect is smaller than expected.
We close with a discussion of our findings.

\section{Model and techniques}
In the following we consider a melt of symmetric $AB$ diblock copolymers confined into a thin film. The chain length is denoted by $N$. One half of the diblock 
consists of $A$ monomers, while the other half consists $B$ monomers. There is a short range repulsion $\chi$ between $A$ and $B$ monomers, which drives the microphase 
separation. The lateral extension of the film is denoted as $L$, while $\Delta_0$ is the distance between the parallel, hard confining surfaces. The monomer density in the 
middle of the film is denoted by $\rho$. 

We employ the bond fluctuation model (BFM) for the Monte Carlo (MC) simulations. In this coarse grained lattice model an effective monomer blocks all eight corners of a unit cube
from further occupancy. We work at a monomer number density of $\rho=1/16$; a value which corresponds to a concentrated solution or melt. Monomers along the chain are connected via bond vectors from the set [2,0,0], [2,1,0], [2,1,1], [2,2,1], [3,0,0], [3,1,0] including all permutations 
and sign combinations. The large number of bond vectors allows 87 distinct bond angles and gives a rather good approximation of continuous space properties. 
Monomers interact via a square well potential which is extended over the nearest 54 lattice sites, which constitute the first neighbor shell in the monomer density
pair correlation function. Monomers of the same species attract each other with strength $-\epsilon$ while a contact in the range of the square well potential between unlike
species increases the energy by an amount $\epsilon$. The surfaces are parallel and impenetrable. An $A$ monomer in the two layers adjacent to the surfaces decreases the
energy by $\epsilon_w$, while a $B$ monomer in this region increases the energy by the same amount. In the MC simulations we use chain length $N=32$, and energy parameters 
$\epsilon=0.1769 k_BT$ and $\epsilon_w=0.1 k_BT$. This corresponds to intermediate segregation $\chi N=30$.
The monomer wall interaction is rather weak; the $A$ component does not wet the surface. The configurations of the polymers are updated via local hopping attempts 
of individual monomers, slithering snake-like movements and exchanges of the identity $A \rightleftharpoons B$ of the two blocks. This allows an efficient relaxation 
of the chain conformation.  One Monte Carlo step consists of 3 slithering snake attempts per chain, 1 local hopping attempt per monomer, and 1 $A \rightleftharpoons B$ 
flip per diblock. 

The self consistent field (SCF) calculations employ the Gaussian chain model. In order to map the bond fluctuation model onto the Gaussian chain model we choose the statistical
segment length $b$ such that $R^2=b^2(N-1)\approx 289$ equals the end-to-end distance of the chains in the Monte Carlo (MC) simulations in the athermal limit ($\epsilon=0$). 
All distances are measured in units of the lattice spacing.  In accord with previous studies we obtain $b=3.05$\cite{OLD}.  Monomers of different species repel each other via a contact 
interaction of strength $\chi$. This Flory Huggins parameter $\chi$ is related to the potential well depth in the MC simulations via $\chi = 2 z_c \epsilon/k_BT$, where $z_c$ 
denotes the effective coordination number\cite{M0}. The latter quantity has been extracted from the intermolecular paircorrelation function to be $z_c=2.65$. Hence, we estimate that 
$\epsilon=0.1769 k_BT$ corresponds to $\chi N=30$ in the Gaussian chain model. This correspondence between the parameters of the bond fluctuation model and the Gaussian chain model 
has proven useful for studying the bulk and interfacial properties of binary homopolymer blends\cite{MREV} and ternary mixtures of two homopolymers and a diblock copolymer\cite{MS1}. The effect of the 
confining surfaces is twofold. On the one hand the monomer density decays continuously from its value $\rho$ in the center of the film to zero in a narrow region of width 
$\Delta_w=0.15 R_e$ close to the surface. On the other hand, $A$ monomers in this surface region are attracted while $B$ monomers are repelled. The strength of this monomer wall 
interaction is chosen such that the surface energy in the MC simulations coincides with the value of the surface free energy in the SCF calculations in the limit that one species
covers the surface completely and packing effects are neglected (cf. paper I). Using a SCF technique of Matsen\cite{MSCF} we have numerically solved the model in mean field approximation. 
Further details of the MC simulations and the SCF calculations can be found in paper I. 

\begin{figure}[htbp]
    \begin{minipage}[t]{160mm}%
       \setlength{\epsfxsize}{10cm}
       \mbox{
        \epsffile{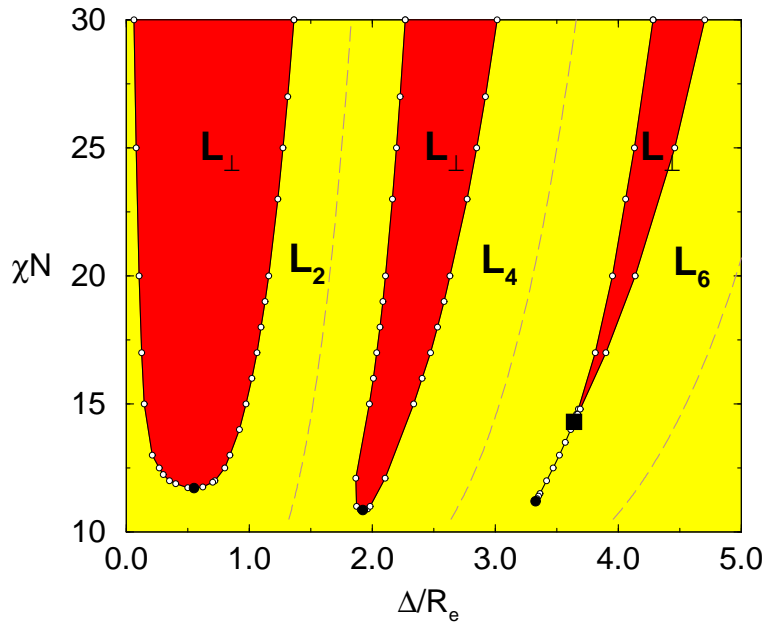}
       }
    \end{minipage}%
    \hfill%
    \begin{minipage}[b]{160mm}%
    \vspace*{1cm}
       \caption{
       \label{fig:phase} Phase diagram of a thin film with symmetric walls $\Lambda_1 N = \Lambda_2 N = 0.2$ as a function of the 
       incompatibility $\chi N$ and the film thickness $\Delta/R_e$.  $L_2$, $L_4$, and $L_6$ denote parallel lamellar phases
       with $2$,$4$, and $6$ $AB$ interfaces, whereas $L_\perp$ denotes the perpendicular lamellar phase. The dashed lines mark
       multiples of the bulk lamellar period. The square denotes the approximate location of the triple point.
       }
    \end{minipage}%
\end{figure}

The phase diagram in the SCF calculations is presented in Fig.\ref{fig:phase}. At high segregation, we observe an alternation of parallel oriented lamellae if the
film thickness $\Delta$ matches a multiple of the bulk period $D_b$, and perpendicular aligned lamellae at intermediate values of the film thickness. This behavior corresponds to the parameters of the
simulation $\chi N = 30$ and $1.71 \leq \Delta/R_e \leq 3.2$. At lower segregation or larger film thickness, the frustration of the parallel lamellar phase due to a mismatch between the
intrinsic length scale of the lamellar order and the confinement decreases, and one encounters direct transitions between parallel lamellar phases. In this case the perpendicular lamellar
phase $L_\perp$ and the adjacent parallel lamellar phase form a triple point. For very small film thickness $\Delta < 3D_b$ we find a second order transition between the perpendicular lamellar 
phase and the disordered phase. Note that there is no transition between the disordered state and the parallel ordered morphologies. Upon increasing the incompatibility the order propagates
gradually from the surfaces into the bulk (cf.\ also Fig.7 in paper I).

\section{Results.}
\begin{figure}[htbp]
\begin{minipage}[t]{160mm}%
   \mbox{
   \setlength{\epsfxsize}{8cm}
   \epsffile{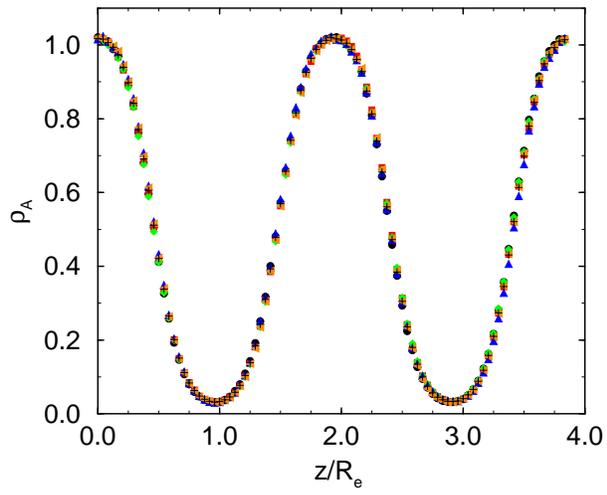}
   }
   \end{minipage}%
  \hfill%
   \begin{minipage}[b]{160mm}%
   \caption{
   \label{fig:plongstat} 
   $A$ density profiles of 6 independent quenches from $\chi=0$ to the $L_\perp$ phase at $\chi N=30$ for film thickness $D=46$.
   }
\end{minipage}%
\end{figure}

Several independent systems are quenched from $\epsilon=0$ to $\epsilon=0.1769 k_BT$. For film thickness $\Delta_0=30$ and $\Delta_0=56$ they  assemble into parallel oriented lamellar phases $L_2$ 
and $L_4$ with two or four $AB$ interfaces, while we find a perpendicular oriented lamellar phase for film thickness $\Delta_0=46$.
The Monte Carlo runs consists of at least $3.8 \cdot 10^6$ Monte Carlo steps. We do not observe transitions between different morphologies within a simulation run and cannot rule out
non-equilibrium effects completely. But the morphologies which are observed agree for all but one system of thickness $\Delta_0=56$ with the predictions of the self-consistent field (SCF) theory.
Moreover, the structures obtained are free of defects and the laterally averaged composition profiles of independent quenches agree to a high accuracy. This is illustrated in 
Fig.\ref{fig:plongstat} for $\Delta_0=46$, where 6 systems have assembled into a perpendicular lamellar phase with a repeat distance $D=1.936 R_e$. This value is about $6\%$ larger than 
the prediction of the SCF theory. The quantitative agreement of the composition profiles of independent systems shows that each system has sampled the fluctuations of the internal interfaces
appropriately.

\subsection{Interfacial fluctuations}
\begin{figure}[htbp]
    \begin{minipage}[t]{160mm}%
       \mbox{
        \setlength{\epsfxsize}{8cm}
        \epsffile{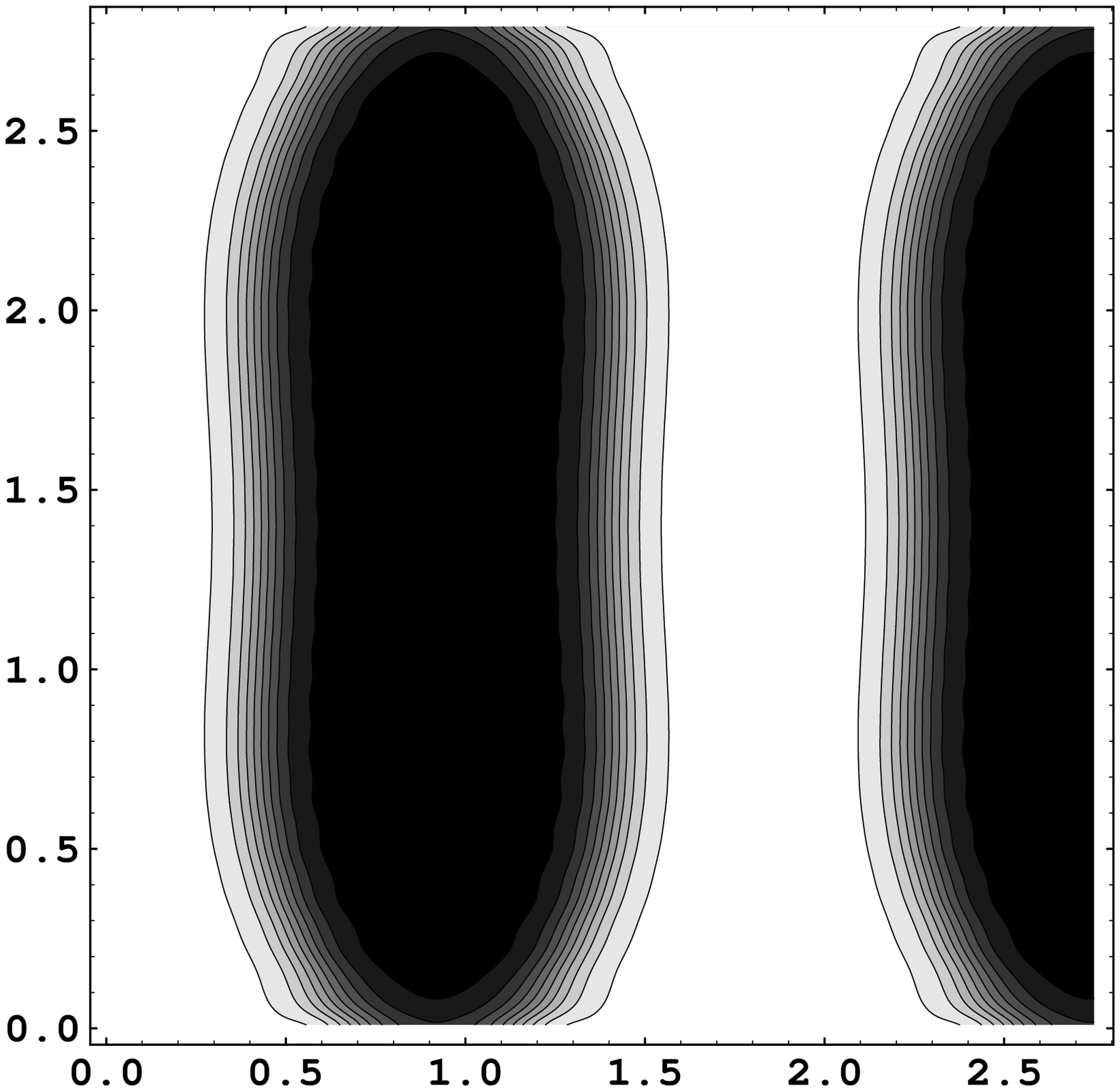}
       }\\
       \vspace*{2cm}
       \mbox{
        \setlength{\epsfxsize}{8cm}
        \epsffile{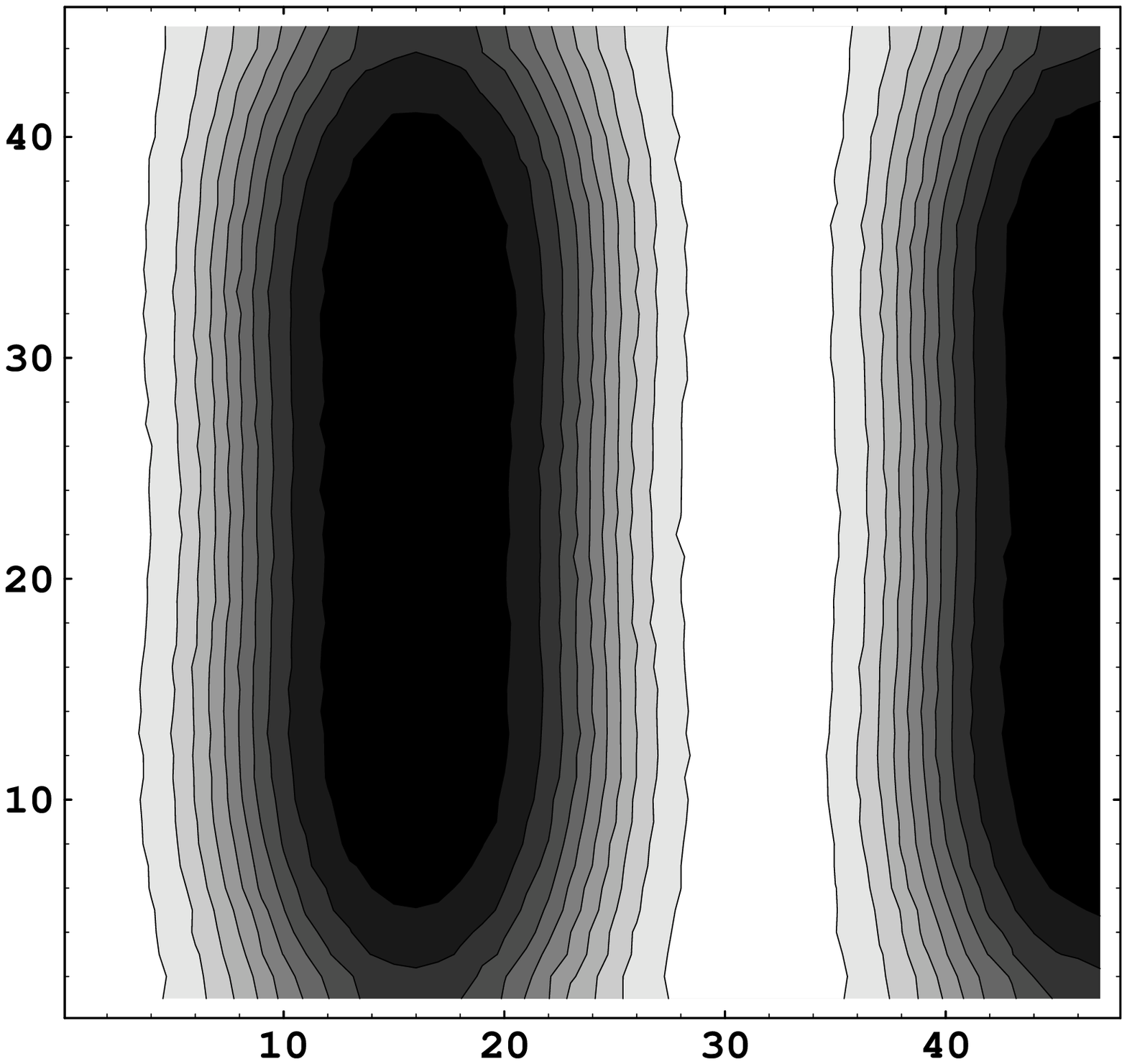}
       }
    \end{minipage}%
    \hfill%
    \begin{minipage}[b]{160mm}%
    \vspace*{1cm}
       \caption{
       \label{fig:LS-SCF-MC} Contour plots of the composition in the perpendicular lamellar phase in a symmetric film $\Delta/R_e=1.83$ and $\Lambda_1 N = \Lambda_2 N = 0.375$
       at $\chi N=30$
       ({\bf a}) SCF theory (lengths are measured in units of the end-to-end distance $R_e$
       ({\bf b}) MC simulations (lengths are measured in units of the lattice spacing/ $R_e=17$ lattice spacings).
       }
    \end{minipage}%
\end{figure}

Fig.\ref{fig:LS-SCF-MC} compares the two dimensional composition profiles in the perpendicular oriented lamellar phase at $\chi N=30$ and $\Delta/R_e=1.83$. The walls ($z$-axis) run horizontally 
on the top and on the bottom of the figure. The $x$ axis across the film is vertical. In each configuration of the MC simulations we have located the instantaneous position of the lamellae
in the $z$ direction, and shifted the $z$ coordinate such that the center of an  $A$ lamella is located at $z=0$. These shifted profiles have been averaged in the $y$ direction and over all configurations.
Hence, the profiles in the MC simulations are broadened by interfacial fluctuations with wavevectors in the $xy$ plane. The  effect of these capillary waves is readily observed:
the interfacial width in the MC simulations is significantly broader than in the SCF calculations. Similar deviations have also been observed in other studies\cite{CAP,Shull}.

Apart from this broadening of the profiles, the MC simulations and the SCF calculations share many subtle details. In both cases the $AB$ interface bends in the vicinity
of the wall as to increase the surface fraction covered with the component of the lower surface free energy. The bending of the interface in the vicinity of the surfaces interferes with the distortion
which originates from the opposite surface and leads to an oscillation of the interfacial position. This has been observed in SCF calculation\cite{Pickett} and is also born out in the MC simulations.
In both cases the interfacial width increases in the vicinity of the surfaces. In the SCF calculations this is due to a reduction of the density at the surface, while in the MC simulation the 
effect stems from the finite interaction range; monomers at the surface have less neighbors to interact with. This gives rise to a negative line tension when the $AB$ interface approaches the 
surface\cite{MSCF}.

In binary blends the interfacial structure can be characterized by the local interfacial position $u(x,y)$, which depends on the lateral coordinates, and the ``intrinsic'' profiles of the order parameter
across the interface. Neglecting the coupling between long wavelength fluctuations of the local interfacial position and the ``intrinsic'' profile, the latter can be calculated as a
profile across an ideally flat interface. The SCF technique calculates these ``intrinsic'' profiles. The fluctuations of the local interfacial position in binary blends
are well describable by the capillary wave Hamiltonian\cite{Buff}:
\begin{equation}
{\cal H}[u({\bf r}_\|)] = \frac{\sigma_{\rm eff}}{2} \int {\rm d}^2{\bf r}_\| \left[ \nabla u \right]^2
\end{equation}
where $\sigma_{\rm eff}$ denotes the effective interfacial tension between the unmixed phases and $u$ denotes the deviation of the local interfacial position from its lateral average. For unconfined 
interfaces in blends the value of the effective interfacial tension agrees well with independent measurements of the free energy costs of an $AB$ interface or SCF calculations\cite{MW}. 
For interfaces in binary blends near a wall the effective interfacial tension extracted from the fluctuation spectrum depends on the distance between the interface and the wall\cite{WET}, and 
the interaction between the interface and the wall imparts a long wavelength cut--off to the spectrum of interfacial fluctuations\cite{Schick}.

\begin{figure}[htbp]
    \begin{minipage}[t]{160mm}%
       \setlength{\epsfxsize}{10cm}
       \mbox{
        \setlength{\epsfxsize}{9cm}
        \hspace*{2cm}\epsffile{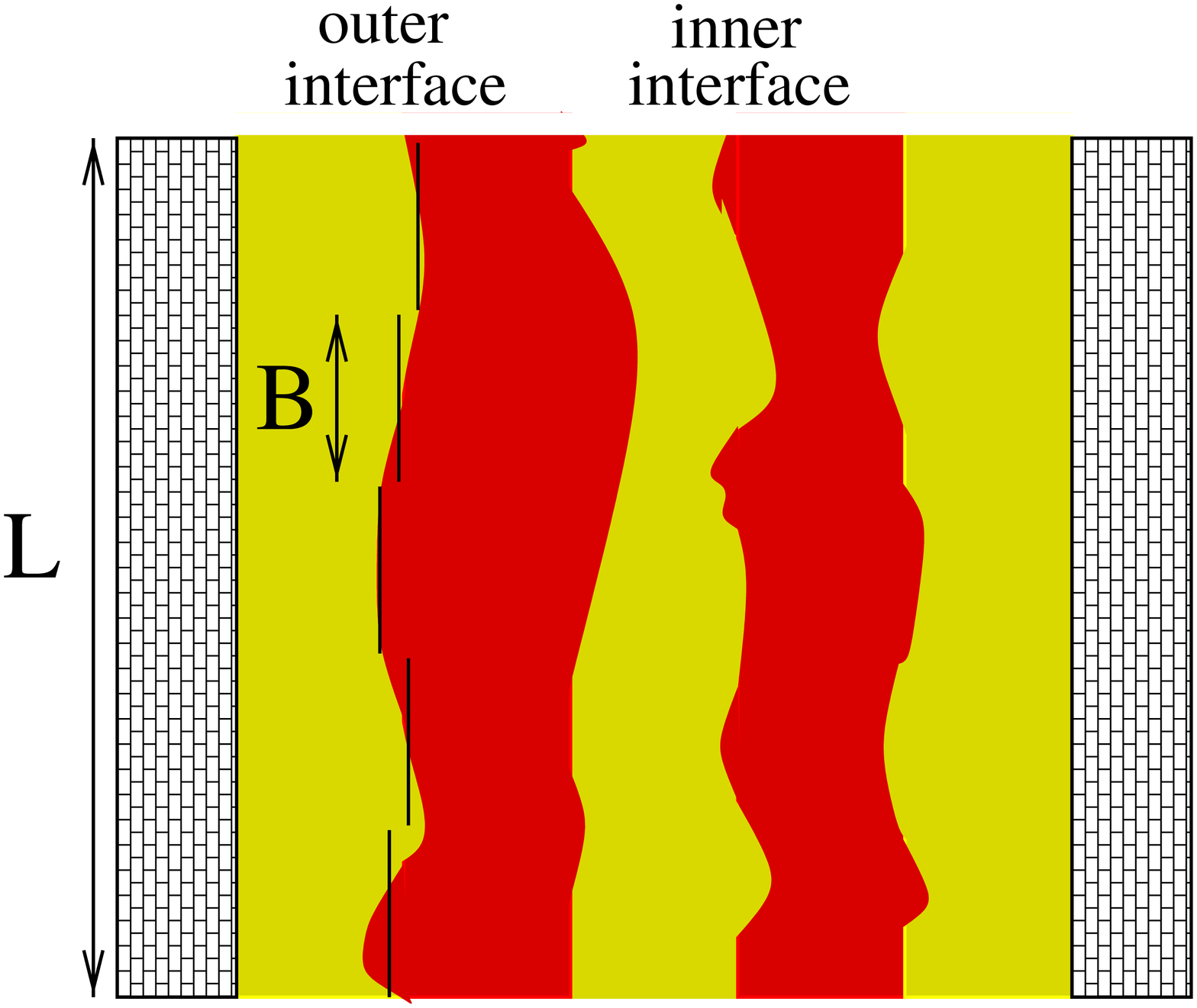}
       }\\
       \vspace*{4cm}
       \mbox{
        \setlength{\epsfxsize}{9cm}
        \epsffile{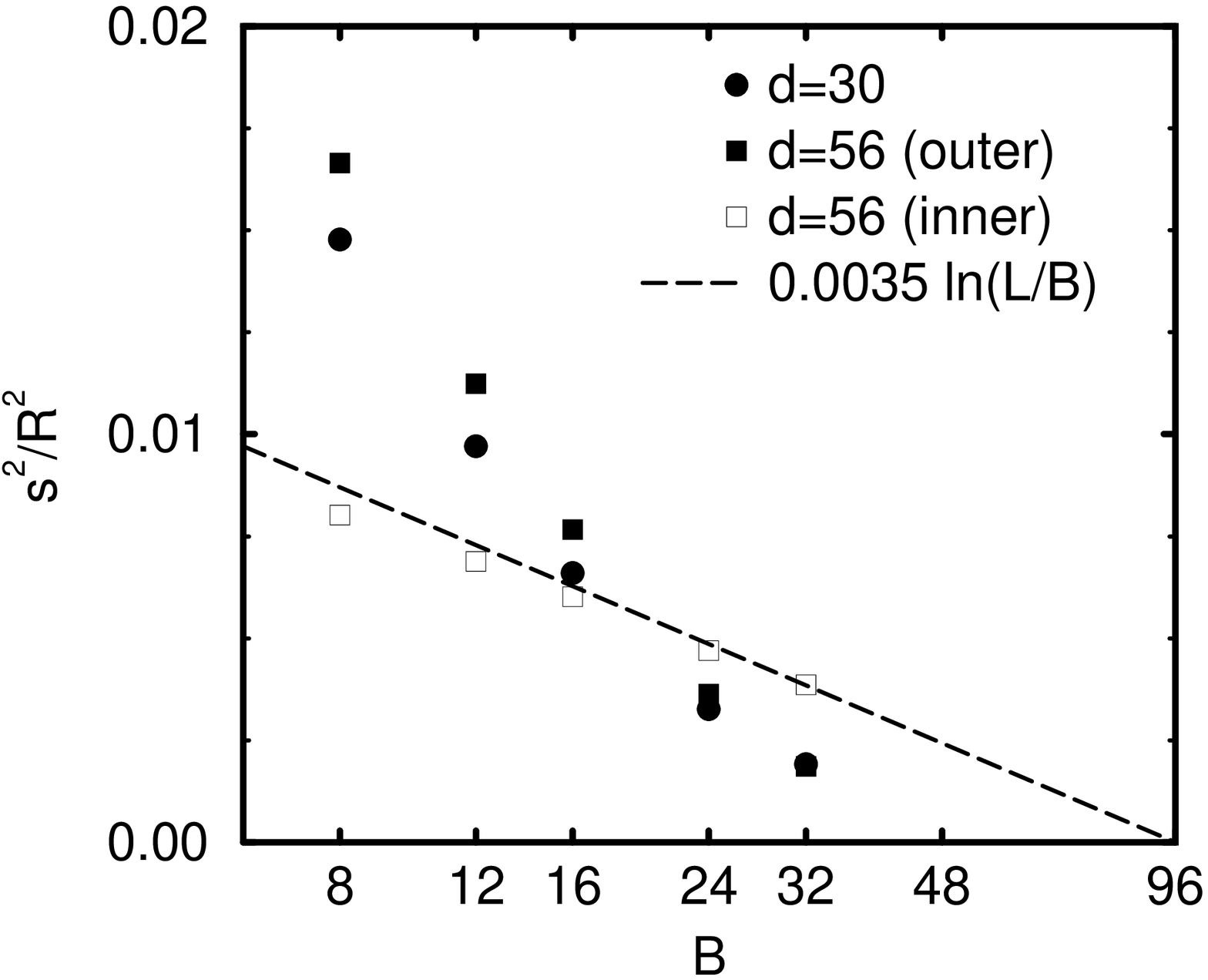}
       }
    \end{minipage}%
    \hfill%
    \begin{minipage}[b]{160mm}%
    \vspace*{1cm}
       \caption{
       \label{fig:cap} Fluctuations of the local interfacial position as a function of the lateral block length $B$:
       ({\bf a}) Schematic illustration of the computational procedure for the $L_4$ phase,({\bf b}) MC results 
       for the parallel lamellar phases $L_2$ and $L_4$ in a symmetric film $\Lambda_1 N = \Lambda_2 N = 0.375$ at $\chi N=30$.
       }
    \end{minipage}%
\end{figure}

We investigate the fluctuations of the local interfacial position in more detail for the parallel lamellar phases $L_2$ and $L_4$. 
In the MC simulations we have measured the local interfacial position via a block analysis\cite{AW1}. This is illustrated schematically in Fig.\ref{fig:cap}({\bf a}). We divide the simulation box into 
lateral columns of size $B \times B$, and determine the local position of the interfaces in each column. The positions are Gaussian distributed around their lateral average across the system 
and $s^2$ denotes the variance of the distribution. Using the capillary wave Hamiltonian one finds:
\begin{equation}
s^2 = \frac{k_BT}{2\pi \sigma_{\rm eff}} \ln \left( \frac{L_{\rm max}}{B}\right)
\label{eqn:s2}
\end{equation}
where $L_{\rm max}$ is the large length scale cut--off for interfacial fluctuations. If the interfaces were unconstrained this cut--off would be set by the lateral 
system extension $L_{\rm max}=L$\cite{AW1}.  In the copolymer system neighboring interfaces interact and this might lead to a noticeable modification of the fluctuation 
spectrum. 

The results of this block analysis is presented in Fig.\ref{fig:cap}({\bf b}), which displays the variance of the interfacial position as a function of the lateral coarse graining size $B$.
Circles refer to the $L_2$ phase while squares correspond to the inner (in the middle of
the film) and outer (close to the surfaces) interfaces of the $L_4$ phase. The fluctuations of the outer interfaces in the $L_4$ phase and the interfaces in the $L_2$ phase are similar.
The spectrum of fluctuations is cut off at about $L_{\rm max} \approx 40$, indicating a rather strong interaction between the interfaces and the wall. The fluctuations for the inner interfaces 
of the $L_4$ phase are well describable by Eq.(\ref{eqn:s2}). The spectrum is cut off by the lateral system size $L_{\rm max}=L=96$. From the slope of the straight line we estimate the effective 
interfacial tension to be: $\sigma_{\rm eff}= 0.157 k_BT$. 

This value can be compared to the interfacial tension of an $AB$ interface in a binary blend\cite{ER} $\sigma_{AB}=\rho b\sqrt{\chi/6}(1-4\ln 2/\chi N) \approx 0.0684 k_BT$. 
Using a similar fluctuation analysis\cite{MS2}, the effective interfacial tension of a single unconfined copolymer bilayer has been measured 
in the framework of the BFM at $\epsilon=0.15 k_BT$ or $\chi N =25$. 
The analysis found that the effective interfacial tension  $\sigma_{\rm eff}=0.03 k_BT$ of a copolymer bilayer with respect to undulations is much smaller than twice the interfacial tension in a binary blend 
$2\sigma_{AB}= 2\times 0.066 k_BT$ at that temperature. The latter finding is in accord with the 
expectation that the free energy cost of the interface in a copolymer melt is smaller than the interfacial tension in a blend. The interfacial tension in a blend is the limiting value at high segregation.

Therefore, our finding for the confined copolymer melt suggests that interfacial fluctuations are strongly suppressed due to the confined geometry and mutual interactions between the interfaces.
Estimating the additional free energy costs of interfacial fluctuations due to chain stretching, Semenov\cite{CAP} argued that equation (\ref{eqn:s2}) (with $\sigma_{\rm eff}=\sigma_{AB}$) is 
appropriate when the spectrum is cut off at length scales larger than the distance between lamellae. Using $\sigma_{AB}$, the width $ \pi w$ of the $AB$ interface as a lower cut--off and the 
periodicity $D$ as an upper cut--off we obtain $s=0.16R_e$. This value is larger than our MC results (cf. next section). Our findings suggest that the effective interfacial tension differs from 
the interfacial tension in a binary polymer blend. Other treatments of waves on the surface of brushes \cite{Fred2,Milner} result in wavevector dependent contributions 
to the free energy of fluctuations of the brush height. Laradji {\em et al.}\cite{ShiNoolandi} have investigated fluctuations around the lamellar solution of the SCF theory 
in the bulk.
Unlike the situation at an interface between partially miscible homopolymers, the eigenmodes of the fluctuations in the lamellar bulk phase are not capillary waves of individual interfaces but
coherent modulations of the whole stack of interfaces. This points to a strong coupling between neighboring lamellae.
Lacking a simple quantitative prediction for the fluctuation spectrum of an ensemble of strongly interacting interfaces in a confined geometry, however, 
we treat the strength of the interfacial fluctuations as an adjustable parameter in the following comparison.

\subsection{Profiles}
\begin{figure}[htbp]
    \begin{minipage}[t]{160mm}%
       \mbox{
        \setlength{\epsfxsize}{8cm}
        \epsffile{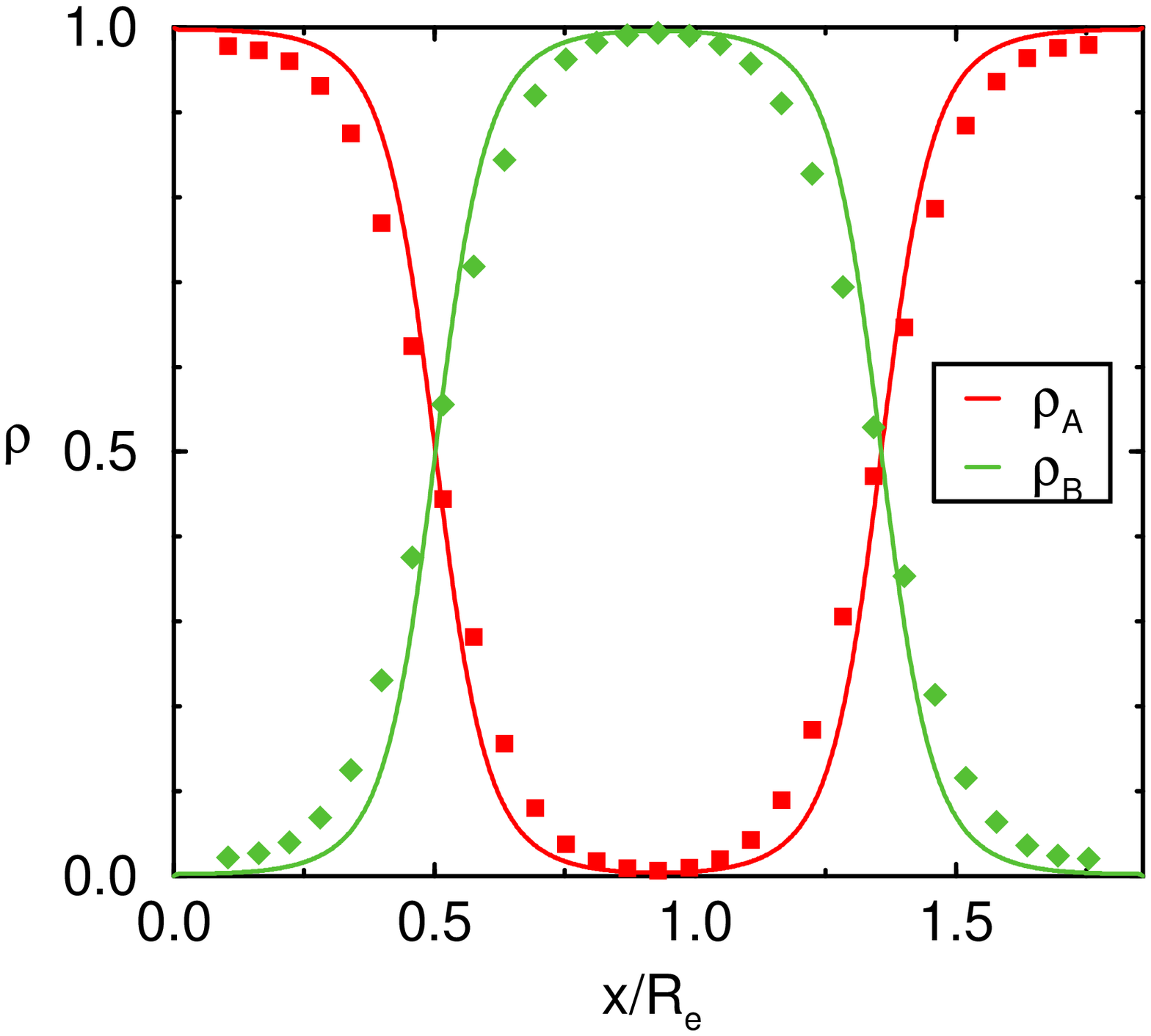}
        \setlength{\epsfxsize}{8cm}
        \epsffile{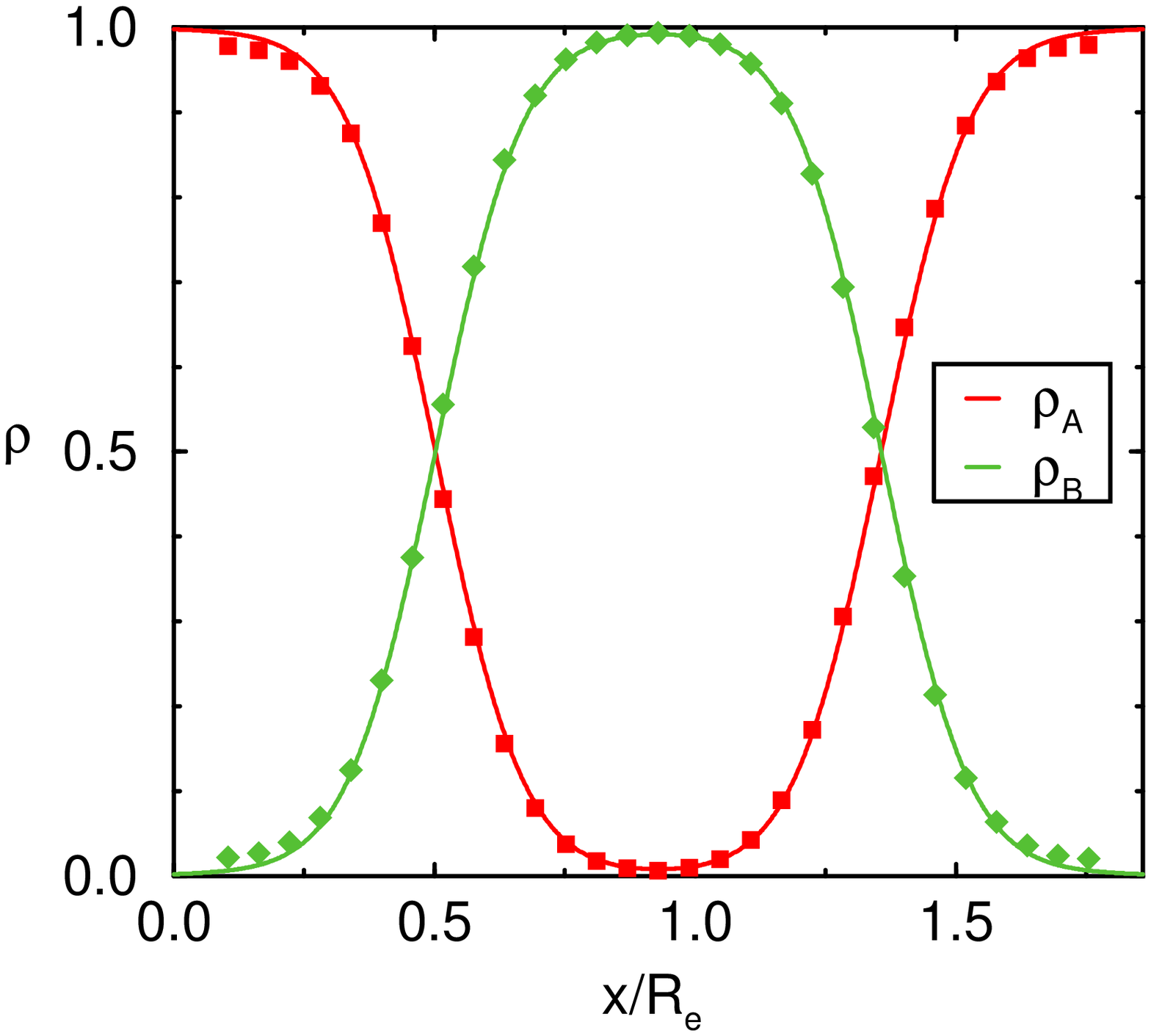}
       }\\
       \mbox{
        \setlength{\epsfxsize}{8cm}
        \epsffile{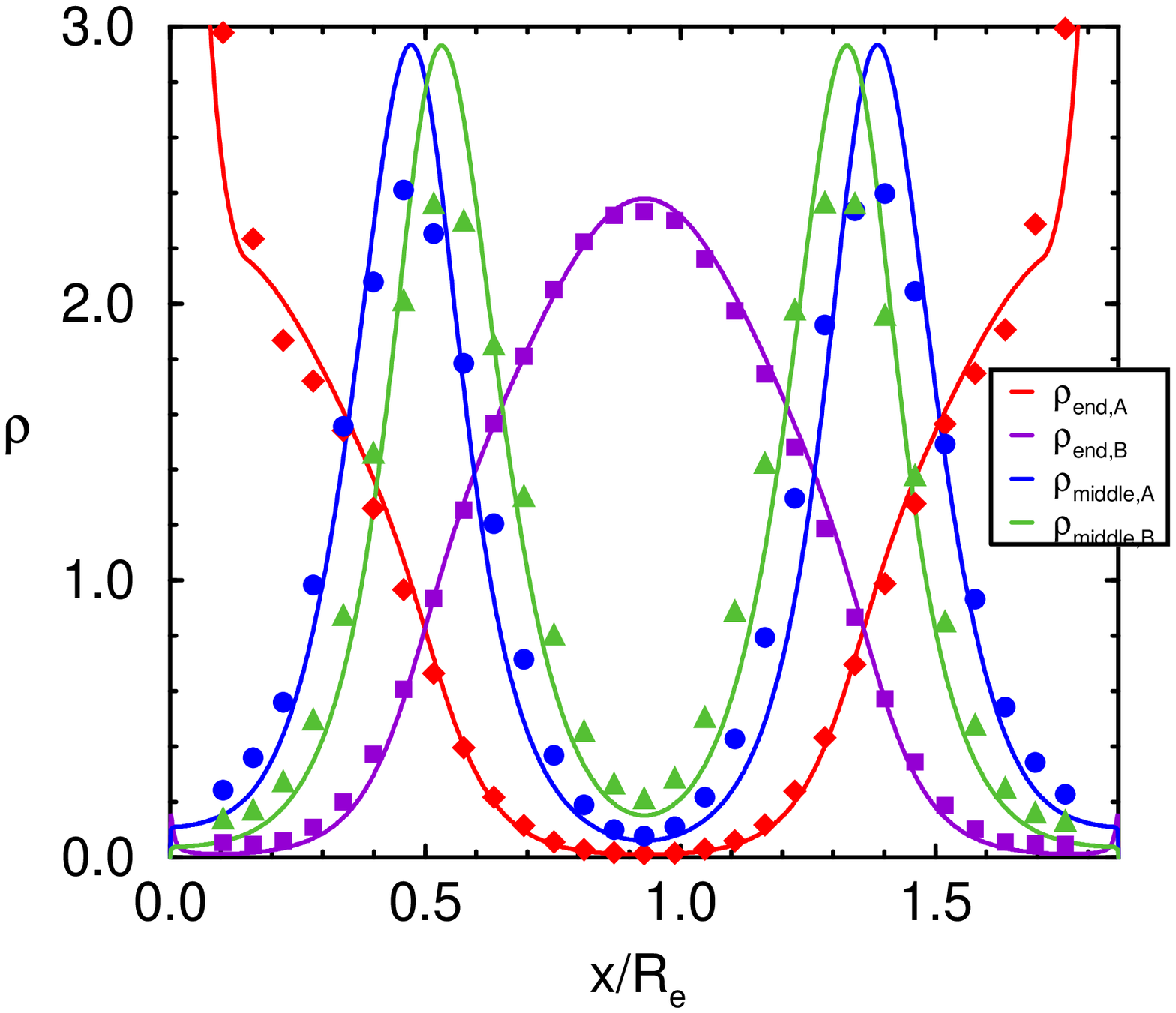}
        \setlength{\epsfxsize}{8cm}
        \epsffile{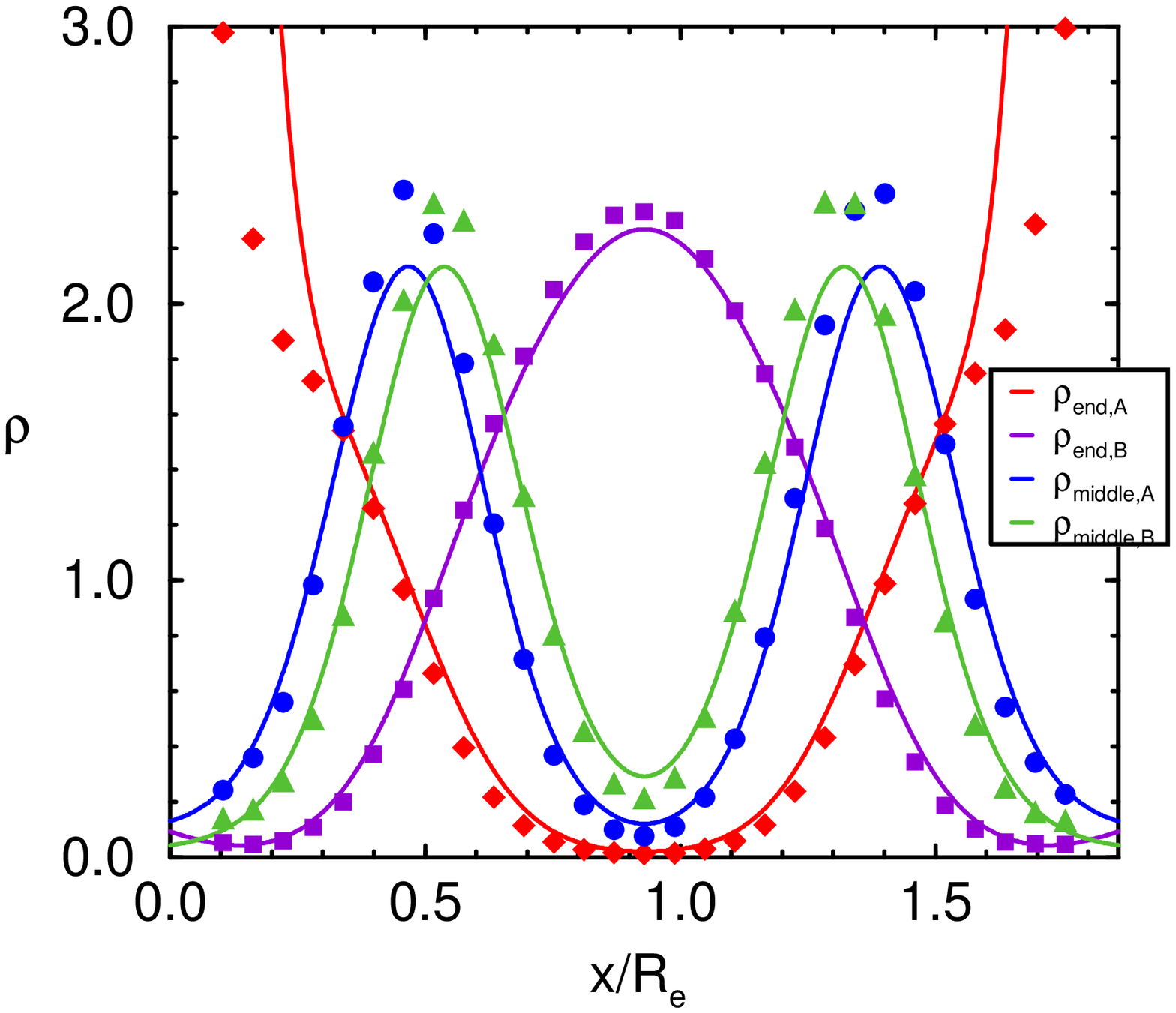}
       }
    \end{minipage}%
    \hfill%
    \begin{minipage}[b]{160mm}%
    \vspace*{1cm}
       \caption{
       \label{fig:L2} Comparison between the results of the SCF theory and the MC simulations:
                 $L_2$ phase in a symmetric film $\Delta/R_e=1.71$ and $\Lambda_1 N = \Lambda_2 N = 0.375$ at $\chi N= 30$.
       ({\bf a}) composition profiles: symbols correspond to the MC results, lines represent the SCF calculations.
       ({\bf b}) composition profiles: SCF profiles broadened with $s=0.0989$.
       ({\bf c}) segmental profiles, raw data: circles and triangles denote the middle monomers of the $A$ block and the $B$ block; diamonds and squares denote the
                 end segments of the $A$ and $B$ block, respectively. 
       ({\bf d}) segmental profiles, SCF profiles broadened with $s=0.0989$. Symbols as in ({\bf c}).
       }
    \end{minipage}%
\end{figure}

The comparison between the composition profiles in the $L_2$ phase in the MC simulations and the SCF calculations is displayed in Fig.\ref{fig:L2}({\bf a}). As discussed above, we find qualitative agreement,
while there are quantitative deviations due to interfacial fluctuations. To account for these effects, we assume that the intrinsic profiles averaged over a small lateral patch of the interface
are describable by the SCF calculations while the local position of the interface is Gaussian distributed. To supplement the SCF results with this additional broadening we convolute the profiles $p$
of any quantity with a Gaussian distribution\cite{CAP,PRE}:
\begin{equation}
p_{\rm cap}(x) \ \int {\rm d} x'\; p_{\rm SCF}(x') \frac{1}{\sqrt{2\pi s^2}}\exp(-[x-x']^2/2s^2)
\end{equation}
We employ the composition profile to adjust the width $s$ of the distribution as to achieve best agreement between the MC simulations and the SCF calculations. We then use the same value of $s$
to convolute profiles of all other quantities. Note that we assume the strength
of the fluctuations to be uniform across the film. Certainly, this is only a rough approximation. As we have observed in Fig.\ref{fig:cap}({\bf b}) the fluctuations of the inner lamellae are larger
than the fluctuations of those close to the film surfaces in the $L_4$ phase. Hence, we expect $s$ to be larger in the middle of the film and smaller at the surfaces. Therefore, by convoluting the
SCF profiles we overestimate the fluctuation effects at the film surfaces.

Fig.\ref{fig:L2}({\bf b}) shows the comparison between the MC simulations and the convoluted profiles of the SCF calculations with $s/R_e=0.1$. We achieve excellent agreement in the middle
of the film and for the interfacial widths. The minor deviations at the surfaces are due to a reduced strength of interfacial fluctuations and the different structure at the surfaces.
The panels ({\bf c}) and ({\bf d}) present the comparison between the MC results and the unconvoluted and convoluted SCF profiles for individual segment densities. Again, the convolution 
improves the agreement between the MC results and the SCF calculations.  The $A$ ends are enriched at both surfaces, while the $B$ ends are located mostly in the middle of the film. The 
maximum at the center is, however, not very pronounced. This indicates that the strong stretching limit is not yet reached for $\chi N=30$ and the two brushes formed by the copolymers largely 
interdigitate. The profiles of the $A$ middle segment (monomer number 16) and the $B$ middle segment (monomer number 17) exhibit a maximum
at the interfaces. In accord with experimental observations, the distribution of the middle segments is narrower than the distribution of the chain ends\cite{RUSSELL}.

\begin{figure}[htbp]
    \begin{minipage}[t]{160mm}%
       \mbox{
        \setlength{\epsfxsize}{8cm}
        \epsffile{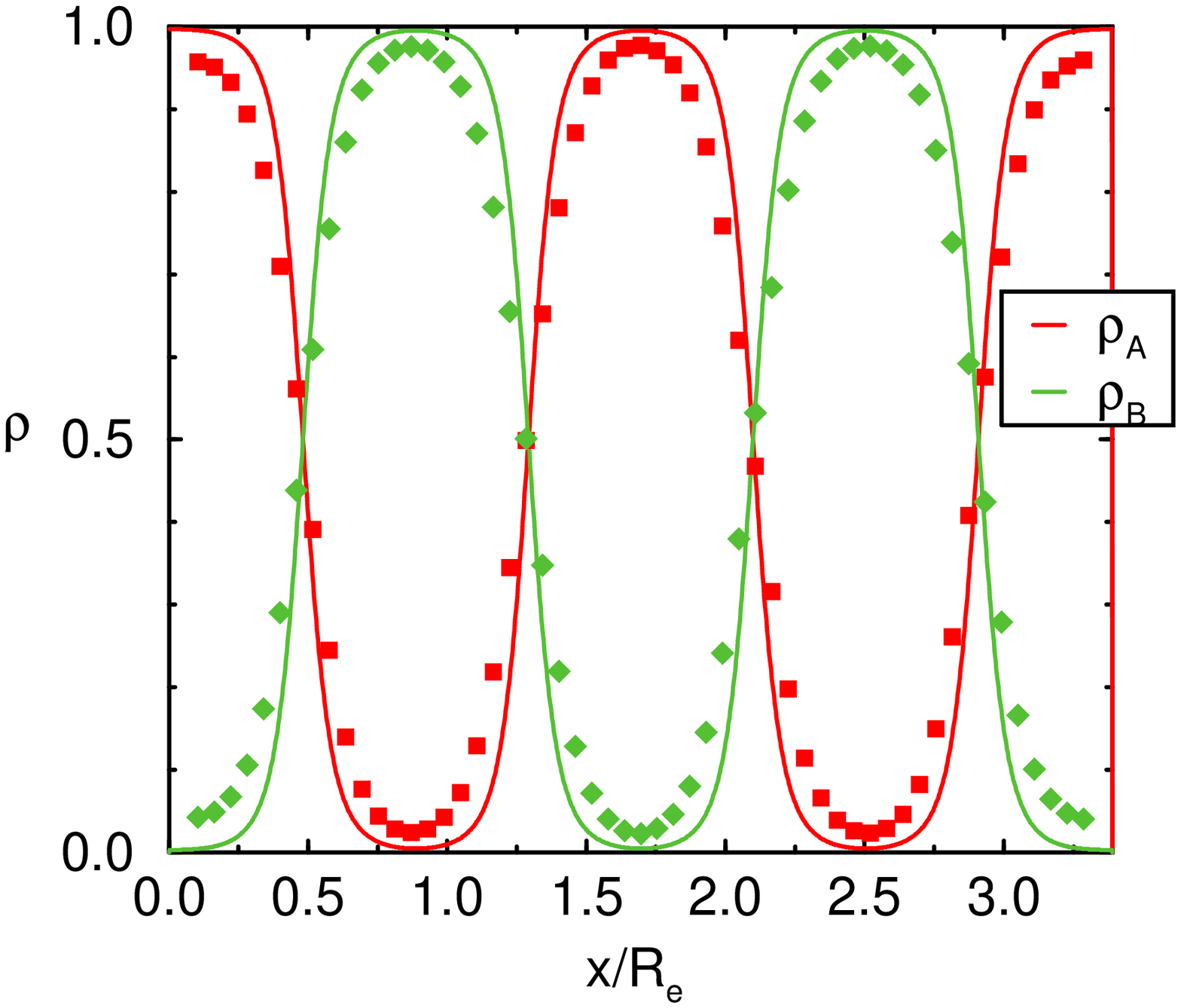}
        \setlength{\epsfxsize}{8cm}
        \epsffile{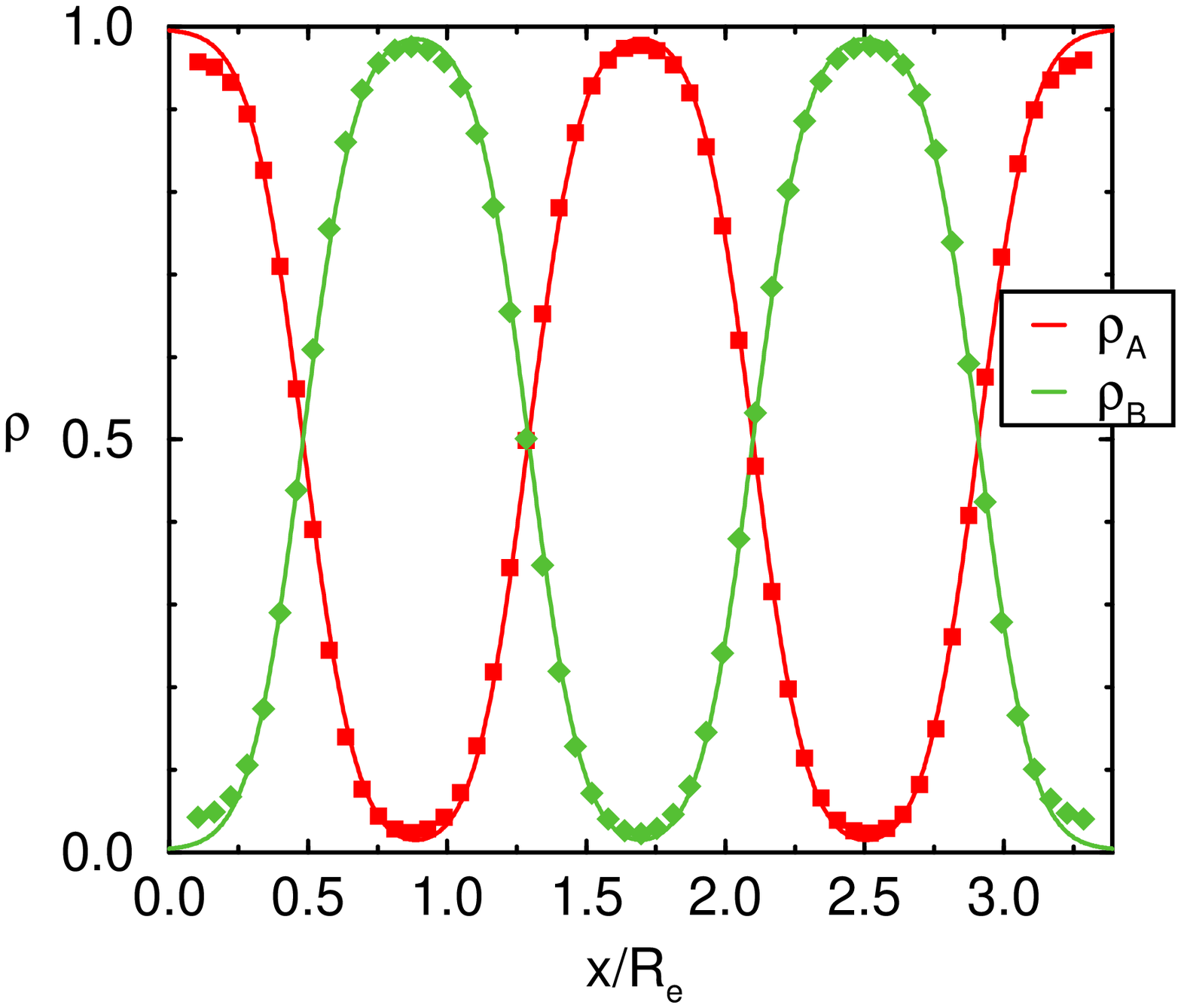}
       }\\
       \mbox{
        \setlength{\epsfxsize}{8cm}
        \epsffile{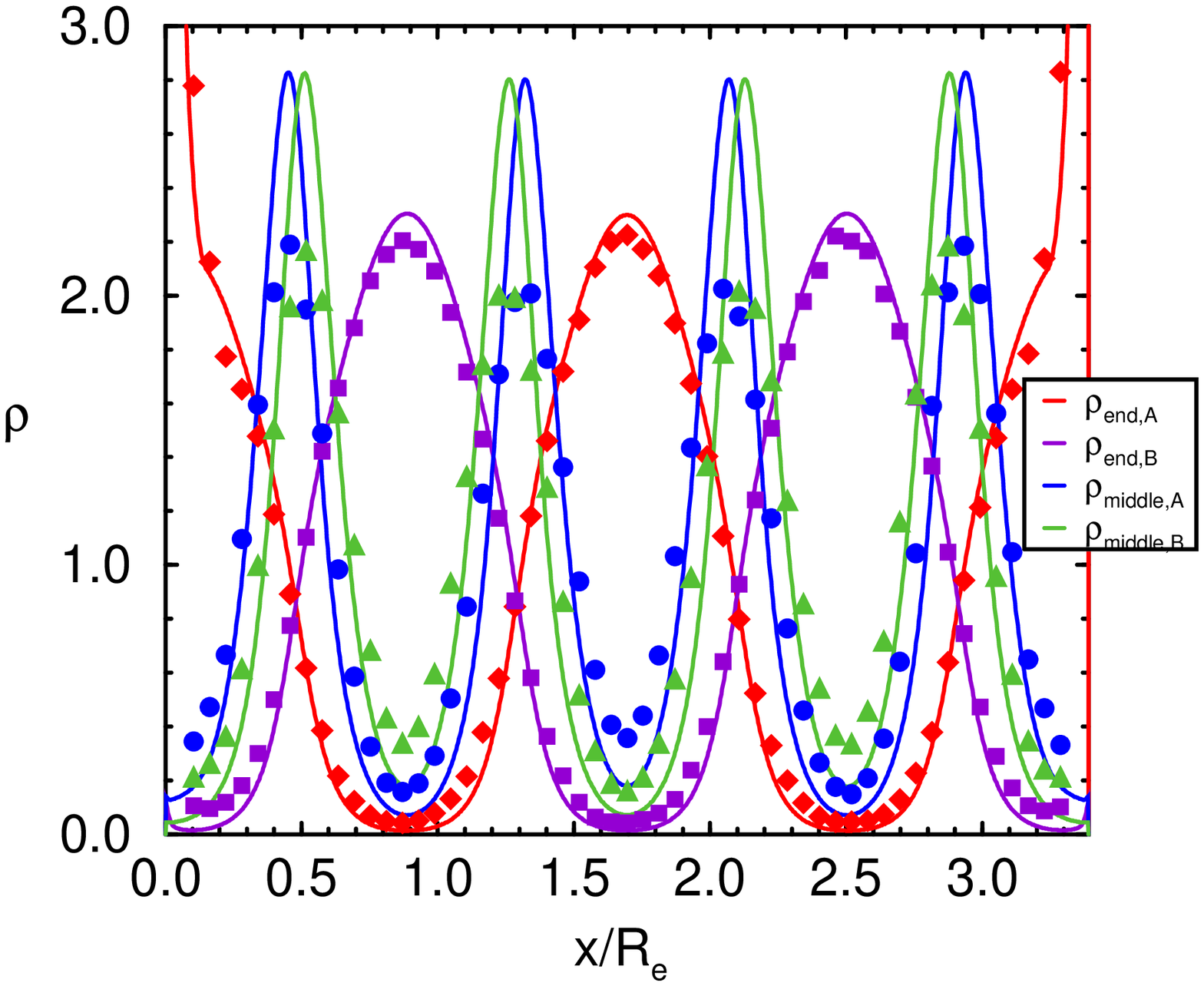}
        \setlength{\epsfxsize}{8cm}
        \epsffile{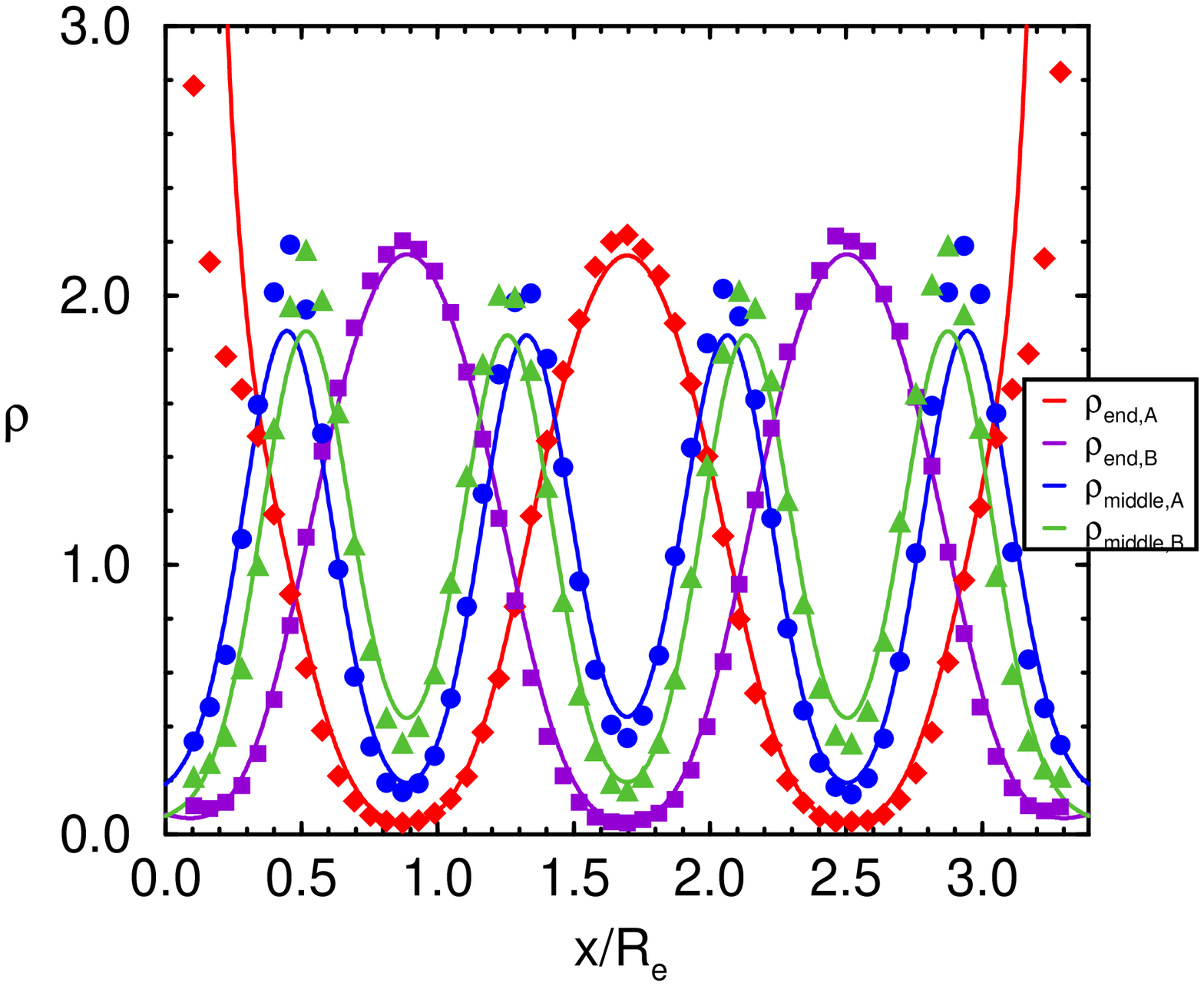}
       }
    \end{minipage}%
    \hfill%
    \begin{minipage}[b]{160mm}%
    \vspace*{1cm}
       \caption{
       \label{fig:L4} Comparison between the results of the SCF theory and the MC simulations:
                 $L_4$ phase in a symmetric film $\Delta/R_e=3.24$ and $\Lambda_1 N = \Lambda_2 N = 0.375$ at $\chi N= 30$.
       ({\bf a}) composition profiles, symbols correspond to the MC results, lines represent the SCF calculations.
       ({\bf b}) composition profiles, SCF profiles broadened with $s=0.12$.
       ({\bf c}) segmental profiles, raw data: circles and triangles denote the middle monomers of the $A$ block and the $B$ block; diamonds and squares denote the
                 end segments of the $A$ and $B$ block, respectively. 
       ({\bf d}) segmental profiles, SCF profiles broadened with $s=0.12$. Symbols as in ({\bf c}).
       }
    \end{minipage}%
\end{figure}

\begin{figure}[htbp]
    \begin{minipage}[t]{160mm}%
       \mbox{
        \setlength{\epsfxsize}{8cm}
        \epsffile{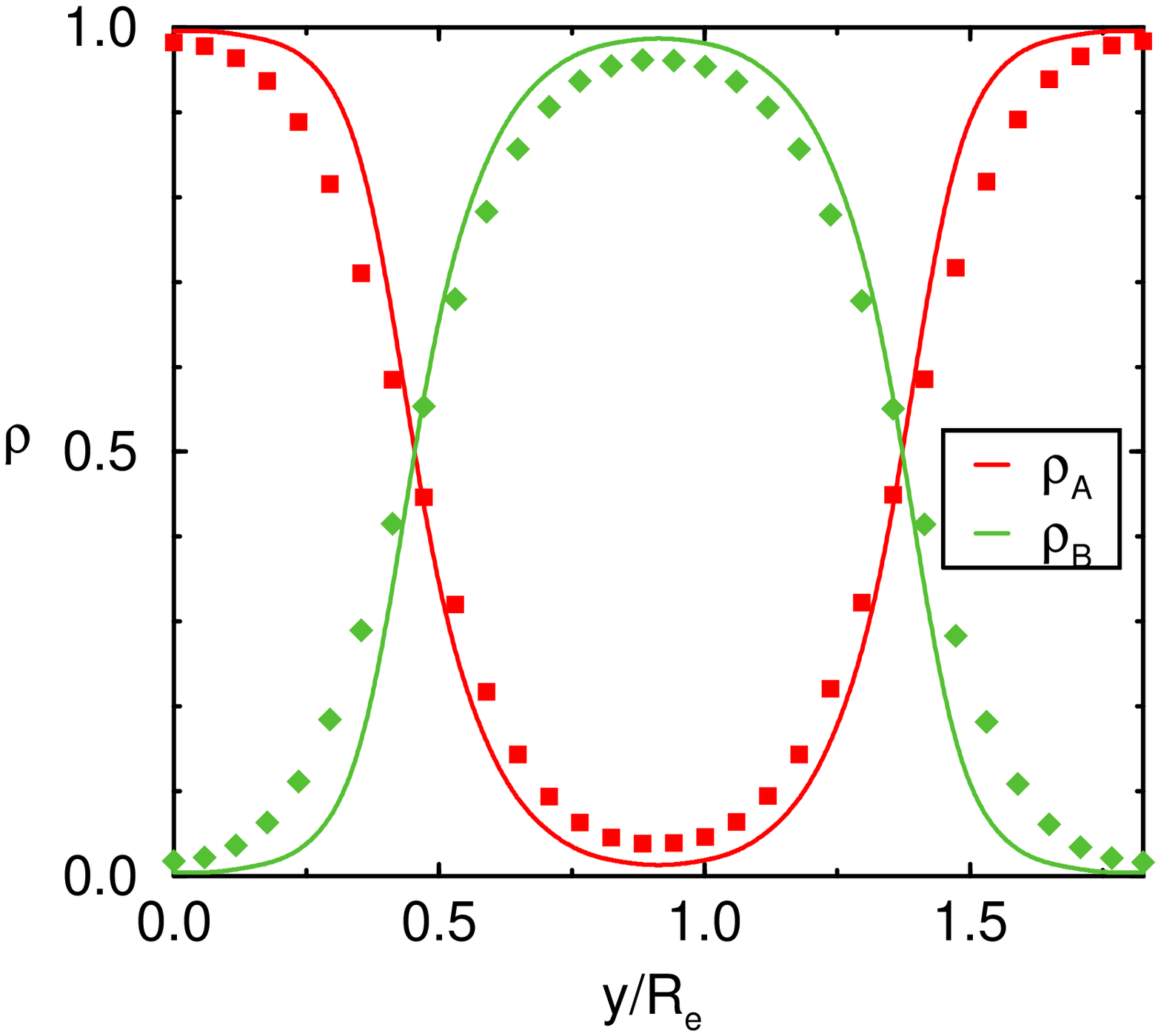}
        \setlength{\epsfxsize}{8cm}
        \epsffile{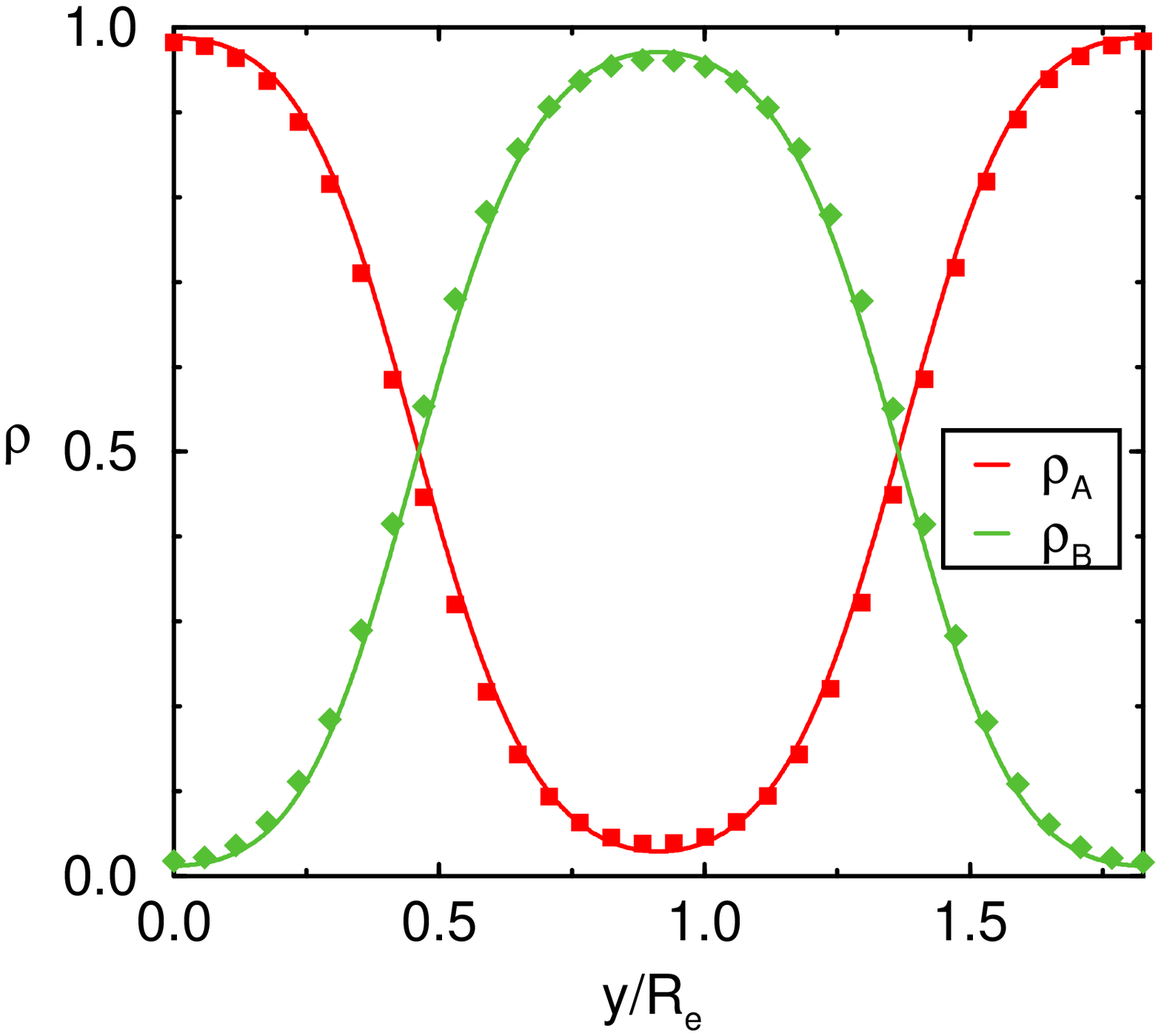}
       }\\
       \mbox{
        \setlength{\epsfxsize}{8cm}
        \epsffile{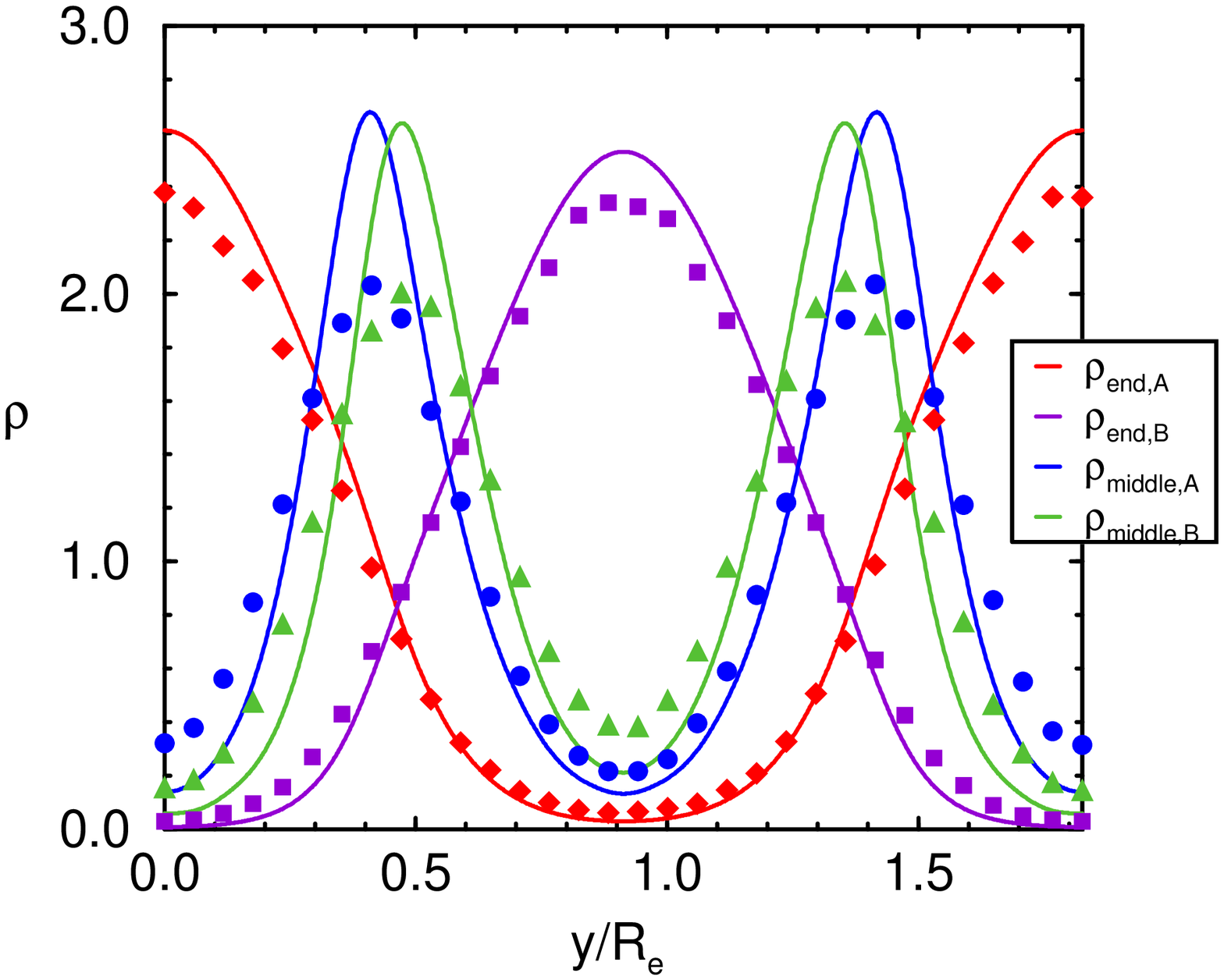}
        \setlength{\epsfxsize}{8cm}
        \epsffile{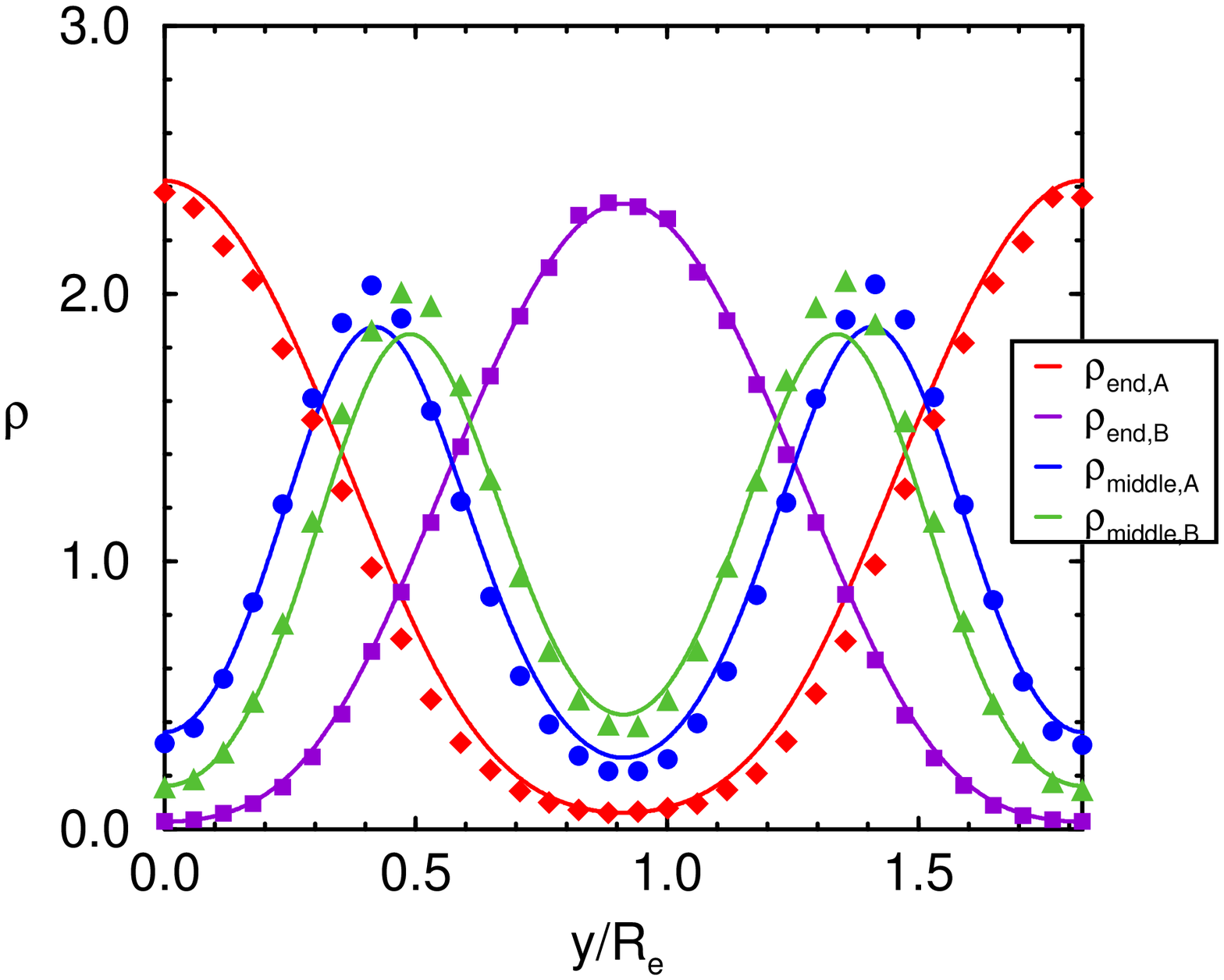}
       }
    \end{minipage}%
    \hfill%
    \begin{minipage}[b]{160mm}%
    \vspace*{1cm}
       \caption{
       \label{fig:Ls} Comparison between the results of the SCF theory and the MC simulations:
                 $L_\perp$ phase in a symmetric film $\Delta/R_e=1.83$ and $\Lambda_1 N = \Lambda_2 N = 0.375$ at $\chi N= 30$.
       ({\bf a}) composition profiles, symbols correspond to the MC results, lines represent the SCF calculations.
       ({\bf b}) composition profiles, SCF profiles broadened with $s=0.127$.
       ({\bf c}) segmental profiles, raw data: circles and triangles denote the middle monomers of the $A$ block and the $B$ block; diamonds and squares denote the
                 end segments of the $A$ and $B$ block, respectively. 
       ({\bf d}) segmental profiles, SCF profiles broadened with $s=0.127$. Symbols as in ({\bf c}).
       }
    \end{minipage}%
\end{figure}

A similar analysis has been performed for the $L_4$ phase and the $L_\perp$ phase. The results are presented in Figs.\ref{fig:L4} and \ref{fig:Ls}, respectively. 
The profiles in the $L_4$ phase are qualitatively similar to those of the $L_2$ phase. In the perpendicular phase $L_\perp$ the profiles run perpendicular to the direction of the $AB$ interfaces of the lamellae
and parallel to the walls. Since the $A$ component is slightly enriched at both surfaces also in the $B$ lamellae, the profile averaged across the film is less segregated in the $B$ lamellae than in the $A$ 
lamellae. This effect is observed in the MC simulations and the SCF profiles.

For each phase we adjusted the strength of the interfacial fluctuations as to match the SCF profiles onto the MC data. For the profiles of all other quantities we used the same value 
of $s$, and thereby improved the agreement between the MC simulations and 
the SCF calculations. This procedure results in $s/R_e=0.1$, $s/R_e=0.12$, and $s/R_e=0.127$ for the $L_2$,$L_4$ and $L_\perp$ phase, respectively.
The increase of the strength of fluctuations is compatible with the intuition that the confined geometry reduces the interfacial fluctuations more in the $L_2$ phase than in the $L_4$ or $L_\perp$ phase.
Using these values of $s$ we can estimate the lateral block size $B_{\rm min}$ on which the MC data agree with the ``intrinsic'' profiles of the SCF calculations according to Fig.\ref{fig:cap}({\bf b}).
This yields the rough estimate $B_{\rm min} \approx 10$.  Semenov suggested this cut--off to be $B_{\rm min}=\pi w$ for binary polymer blends and a recent
MC study\cite{PRE} in the framework of the BFM found $B_{\rm min}=1.2 \pi w(1-3.1/\chi N)$. Using $w=b/\sqrt{6\chi}$, we obtain $B_{\rm min}\approx 4.5$. This is again an indication that the 
interfacial fluctuations are not well describable with the capillary wave Hamiltonian in the confined copolymer system. One possible explanation for a larger value of the cut--off $B_{\rm min}$ 
is, e.g., a bending rigidity of the lamellae.

\section{ Summary. }
We have presented SCF calculations and MC simulations for symmetric diblock copolymers confined into a thin film. Both surfaces attract the same component of the diblock via a short range
potential. We have quenched several independent systems from the athermal state to $\chi N=30$, and we have compared the results of the MC simulations in the framework of the BFM with
SCF calculations in the Gaussian chain model. We find qualitative agreement between the MC simulations and the SCF calculations. In particular, we observed the $L_2$, $L_\perp$ and $L_4$ 
phases as predicted by the SCF calculations (cf. paper I). We have used these configurations to investigate the detailed structure in the thin film geometry.

In the simulations we find evidence for a broadening of the profiles due to interfacial fluctuations. However, the spectrum of interfacial fluctuations is not well describable by the capillary 
wave Hamiltonian. The effective interfacial tension is higher than the interfacial tension in a binary blend of $A$ and $B$ homopolymers, and the spectrum is cut---off at large wave lengths. 
A similar increase of the effective interfacial tension of an interface in a binary polymer blend in the vicinity of a wall has been observed\cite{WET}, however, the effect is more pronounced in
the copolymer system.  The confinement and the mutual interactions between neighboring lamellae strongly suppress interfacial fluctuations.

To mimic the effect of interfacial fluctuations we convolute the  SCF profiles with a Gaussian. The strength of the fluctuations is treated as a free parameter for each film thickness, however, 
we use the same value for profiles of different quantities. Taking account of interfacial fluctuations, we find almost quantitative agreement between the MC results and the SCF profiles for the 
composition, the density of chain ends and middle segments.  In agreement with experiments\cite{RUSSELL} both MC simulations and SCF calculations show that the middle segements of the 
copolymer are stronger localized at the $AB$ interface than the ends of the diblock in the middle of the domains. The good agreement between MC simulations and SCF calculations suggests that
composition fluctuations play only a minor role at intermediate segregation. Moreover, the local fluid structure of the bond fluctuation model does not have a large influence on the profiles. 
The latter finding is rather unexpected, because the interactions on the monomer scale $\chi=30/32$ are large and the interfacial width between the $A$ and $B$ domains is not much larger than the 
microscopic length of the model.

\subsection*{Acknowledgment}
It is a great pleasure to thank P.K. Janert, F. Schmid, and M.W. Matsen for valuable discussions/correspondence.
We acknowledge generous access to the CRAY T3E at the HLR Stuttgart and HLRZ J{\"u}lich, as well as access to the CONVEX SPP at the computing center in Mainz.
Financial support was provided by the DFG under grant Bi314/17.


\begin{thebibliography}{99}
\bibitem{Nakanishi}  Fisher M. E.; Nakanishi H. J.Chem.Phys. {\bf 1981}, { 75}, 5857.
                     Nakanishi H.; Fisher M. E. J.Chem.Phys. {\bf 1983}, { 78}, 3279.
\bibitem{Russell} T.P Russell, Current Opinion in Colloid \& Interfacial Science {\bf 1}, 107 (1996);
                  K. Binder, Adv.Pol.Sci {\bf 138},1 (1999).
\bibitem{MSCF} M.W. Matsen, J.Chem.Phys. {\bf 106}, 7781 (1997);
               M.W. Matsen Current Opinion in Colloid and Interfacial Science {\bf 3}, 40 (1998).
\bibitem{C_SULL}  K.R Shull, A.M. Mayes, and T.P. Russell, Macromolecules {\bf 26}, 3929 (1993).
\bibitem{C_MATSEN} N. Koneripalli, R. Levicky, F.S. Bates, M.W. Matsen, S.K. Satija, J. Ankner, and H. Kaiser, Macromolecules {\bf 31}, 3498 (1998).
\bibitem{PAPER1}  T. Geisinger, M. M{\"u}ller, and K. Binder, preceding paper.
\bibitem{Fredrickson} G.H. Fredrickson and E. Helfand, J.Chem.Phys. {\bf 87}, 697 (1987).
\bibitem{David}  E.F. David and K.S. Schweizer, J.Chem.Soc.Faraday Trans. {\bf 91}, 2411 (1995).
\bibitem{Fried} H. Fried and K. Binder, Euro.Phys.Lett. {\bf 16}, 237 (1991); J.Chem.Phys. {\bf 94}, 8349 (1991); Macromolecules {\bf 26}, 6878 (1993).
\bibitem{Maurer} W.W. Maurer, F.S. Bates, T.P. Lodge, K. Almdal, K. Mortensen, and G.H. Fredrickson, J.Chem.Phys. {\bf 108}, 2989 (1998).
\bibitem{Stamm1} K. Almdal, J.H. Rosedale, F.S. Bates, G.D. Wignall, and G.H. Fredrickson, Phys.Rev.Lett. {\bf 65}, 1112 (1990);\\
                 J.H. Rosedale, F.S. Bates, K. Almdal, K. Mortensen, and G.D. Wignall, Macromolecules {\bf 28}, 1429 (1995).
\bibitem{Stamm}  V.T. Bartels, M. Stamm, V. Abetz, and K. Mortensen, Euro.Phys.Lett. {\bf 31}, 81 (1995).
\bibitem{Vilgis} T.A. Vilgis, A. Weyersberger, and M.G. Brereton, Phys.Rev. {\bf E 49}, 3031 (1994).
\bibitem{MW}     M. M\"uller and A. Werner, J.Chem.Phys. {\bf 107}, 10764 (1997).
\bibitem{Nath1}   S.K. Nath, J.D. McCoy, J.G. Curro, and R.S. Saunders, J.Chem.Phys. {\bf 106}, 1950 (1997).
\bibitem{Nath2}   J.D. McCoy, S.K. Nath, J.G. Curro, and R.S. Saunders, J.Chem.Phys. {\bf 108}, 3023 (1998).
\bibitem{Pickett} G.T. Pickett and A.C. Balazs, Macromolecules {\bf 30}, 3097 (1997).
\bibitem{Sommer}  A. Hoffmann, J.U. Sommer, and A. Blumen, J.Chem.Phys. {\bf 107}, 7559 (1997).
\bibitem{OLD} F. Schmid and M. M{\"u}ller, Macromolecules {\bf 28}, 8639 (1995);\\
              M. M{\"u}ller, K. Binder, and W. Oed, J.Chem.Soc. Faraday Trans. 91, 2369 (1995).
\bibitem{M0}  M. M{\"u}ller and K. Binder, Macromolecules {\bf 28}, 1825 (1995).
\bibitem{MREV} for a review see M. M{\"u}ller, Macromolecular Theory and Simulation (in press).
\bibitem{MS1}   M. M\"uller and M. Schick, J.Chem.Phys. {\bf 105}, 8885 (1996).
\bibitem{CAP} A.N. Semenov, Macromolecules {\bf 26},  6617 (1993); Macromolecules {\bf 27}, 2732 (1994).
\bibitem{Shull} K.R. Shull, A.M. Mayes, and T.P Russell, Macromolecules {\bf 26}, 3929 (1993).
\bibitem{Buff} F.P. Buff, R.A. Lovett, and F.H. Stillinger, Phys.Rev.Lett. {\bf 15}, 621 (1965).
\bibitem{WET} M. M\"uller and K. Binder, Macromolecules {\bf 31}, 8323 (1998).
\bibitem{Schick} M. Schick, Les Houches lectures on ``Liquids at interfaces'' {\bf 1990},
		     Elsevier Science Publishers B.V.
\bibitem{AW1} A. Werner, F. Schmid, M{\"u}ller, and K. Binder, J.Chem.Phys. {\bf 107} 8175 (1997).
\bibitem{ER}  A.V. Ermoshkin and A.N Semenov, Macromolecules {\bf 29}, 6294 (1996).
	      A.N. Semenov, J.Phys.II {\bf 6}, 1759 (1997).
\bibitem{MS2} M. M\"uller and M. Schick, J.Chem.Phys. {\bf 105}, 8282 (1996).
\bibitem{Fred2}   G.H. Fredrickson, A. Ajdari, L. Leibler, and J.-P. Carton, Macromolecules {\bf 25}, 2882 (1992).
\bibitem{Milner}  H.-W. Xi and S.T. Milner, Macromolecules {\bf 29}, 4772 (1996).
\bibitem{ShiNoolandi} M. Laradji, A.-C. Shi, J. Noolandi, and R.C. Desai, Macromolecules {\bf 30}, 3242 (1997);
                      M. Laradji, A.-C. Shi,  R.C. Desai, and J. Noolandi, Phys.Rev.Lett. {\bf 78}, 2577 (1997);
		      C. Yeung, A.-C. Shi, J. Noolandi, and R.C. Desai, Macromol.Theory Simul. {\bf 5}, 291 (1996).
\bibitem{PRE} A.Werner, F. Schmid, M. M{\"u}ller, and K. Binder, Phys.Rev. {\bf E 59}, 728 (1999).
\bibitem{RUSSELL}  A.M. Mayes, R.D. Johnson, T.P. Russell, S.D. Smith, S.K. Satija, and C.F. Majkrzak, Macromolecules {\bf 26}, 1047 (1993).




\end{thebibliography}
\end{document}